\documentclass{elsart}
\usepackage{amssymb}
\usepackage{graphicx}
\usepackage{amsmath}

\input{tcilatex}

\begin{document}

\begin{frontmatter}
\title{Regge theory and statistical mechanics}
\author{F. Canfora},
\author{L. Parisi},
\author{G. Vilasi}
\address
{Istituto Nazionale di Fisica Nucleare, Sezione di Napoli, GC di Salerno\\
Dipartimento di Fisica ''E.R.Caianiello'', Universit\`{a} di
Salerno\\ Via S.Allende, 84081 Baronissi (Salerno), Italy}

\begin{abstract}
An interesting connection between the Regge theory of scattering,
the Veneziano amplitude, the Lee-Yang theorems in statistical
mechanics and nonextensive Renyi entropy is addressed. In this
scheme the standard entropy and the Renyi entropy appear to be
different limits of a unique mathematical object. This framework
sheds light on the physical origin of nonextensivity. A non
trivial application to spin glass theory is shortly outlined.
\end{abstract}


\end{frontmatter}

\section*{Introduction}

Quantum information theory is one of the hottest topic in physics because of
its theoretical and phenomenological interest. One of the main tools needed
to analyze the open problems in this fascinating field is the concept of
entropy which, by the way, plays a key role in all theoretical as well as
experimental fields. The standard point of view, which is able to describe a
huge amount of observations, is the standard \textit{BGNS}\footnote{\textit{%
BGNS} stands for Boltzmann-Gibbs-Von Neumann-Shannon.} entropy which is 
\begin{equation*}
S\left( \left\{ p_{i}\right\} \right) =-\dsum\limits_{i}p_{i}\ln p_{i}
\end{equation*}%
where $\left\{ p_{i}\right\} $ is a distribution of probability%
\begin{eqnarray}
\left. \left\{ p_{i}\right\} \quad i=1,..,N\right\vert \quad p_{i} &\geq
&0\quad \forall i,  \label{d1} \\
\dsum\limits_{i}p_{i} &=&1.  \label{d2}
\end{eqnarray}%
From the above definition, the equilibrium thermodynamics (both classical
and quantum) follows, as usual, by using well known constrained
minimizations procedures. On the other hand, there is an increasing amount
of data indicating that, in certain circumstances (such as multiparticles
hadronic systems, fractional diffusion processes, multifractal systems,
coding theory and cryptography, chaotic dynamical systems,
quantum-informations uncertainty relations and so on; for an updated list of
references see \cite{Re61}) a different notion of entropy is more suitable.
In the above mentioned cases the nonextensive \textit{Renyi} entropy \cite%
{Re61} appears to be the right one.

Given a distribution of probability as in Eqs. (\ref{d1}) and (\ref{d2}),
the \textit{Renyi} entropy $S_{R}^{b}$ of index $b$ is defined as%
\begin{equation}
S_{R}^{b}(\left\{ p_{i}\right\} )=\log _{2}\left[ \left(
\dsum\limits_{i}p_{i}^{b}\right) ^{\frac{1}{1-b}}\right] .  \label{rentro}
\end{equation}%
It reduces to the \textit{BGNS} entropy when $b\rightarrow 1$.

Besides the Renyi entropy, also the nonextensive \textit{Tsallis} entropy 
\cite{Tsa88}\ is able to describe many phenomena (such as systems with
long-range interactions) in which the standard extensive entropy seems to
fail. For these reasons it would be important to shed more light on the
conceptual relations between these different notions of entropy. In
particular, it will be shown that the fact that the Renyi entropy (as well
as the Tsallis entropy) approaches the standard one when a suitable
parameter goes to $1$ is not at all the end of the story. One of the main
problems affecting these two notions of entropy is the following: even if
many phenomena are very well described by nonextensive entropies, the index
of nonextensivity changes from phenomenon to phenomenon. It is not clear the
physical information hidden in the nonextensivity index and, therefore, it
is not clear the mechanism which leads to such a "non universality". The
main scope of the paper is to propose an interpretation of the
nonextensivity index able to shed light on these questions.

Indeed, there is a very rich analytical structure connecting these different
entropies which could provide many phenomena with a new understanding (such
as the meaning of the parameter $b$ present in the Renyi entropy of which,
at a present time, it is available only a phenomenological interpretation as
a parameter measuring the nonextensivity).

Quite surprisingly, the mathematical framework which underlies all these
entropies is the Regge theory of scattering \cite{Re1,Re2} when applied to
tree diagrams in string theory. To begin with, let us recall the expression
of the Veneziano amplitude \cite{Ve1} $A_{V}$ 
\begin{eqnarray}
A_{V}(t,s,u) &=&c_{1}\left[ B(-\alpha _{0}(s),-\alpha _{0}(t))+B(-\alpha
_{0}(s),-\alpha _{0}(u))\right.  \label{vene1} \\
&&\left. +B(-\alpha _{0}(t),-\alpha _{0}(u))\right]  \notag \\
B(-\alpha _{0}(x),-\alpha _{0}(y)) &=&\frac{\Gamma (-\alpha ^{\prime
}x-1)\Gamma (-\alpha ^{\prime }y-1)}{\Gamma (-\alpha ^{\prime }x-\alpha
^{\prime }y-2)},\quad  \notag \\
s+t+u &=&N,\quad s,t,u\geq 0  \notag
\end{eqnarray}
where $B$ is the Euler \textit{beta function}, $s$, $t$ and $u$ are the
Mandelstam variables, $\Gamma $ is the Euler gamma function and $c_{1}$, $%
\alpha ^{\prime }$ and $N$ are constants. The beautiful analytical structure
of the Veneziano amplitude, which is based on the Regge theory of
scattering, describes quite well certain features of strong interactions
and, eventually, it has been recognized his string-theoretical origin: it
can be derived by computing the vacuum expectation value of four tachyonic
operators on the disk. As far as the above themes are concerned, two limits
are interesting: the \textit{Regge limit} and the \textit{hard scattering
limit}.

The Regge limit is 
\begin{eqnarray}
\left\vert s\right\vert &\rightarrow &\infty ,\quad t\quad fixed  \notag \\
A_{V} &\thickapprox &c_{1}s^{\alpha _{0}(t)}\Gamma (-\alpha _{0}(t))\approx
c_{1}\exp \left[ \ln \left( s^{\alpha _{0}(t)}\right) \right] \Gamma
(-\alpha _{0}(t)),  \label{rl2} \\
\quad \alpha _{0}(t) &=&\alpha ^{\prime }t+1;  \notag
\end{eqnarray}
the above Regge behavior is valid in all the complex $s-$plane as long as $s$
is far away the real axis.

The hard scattering limit is 
\begin{eqnarray}
\left\vert s\right\vert  &\rightarrow &\infty ,\quad t/s\quad fixed
\label{hs1} \\
A_{V} &\thickapprox &c_{1}\exp \left[ -\alpha ^{\prime }\left( t\ln t+s\ln
s+u\ln u\right) \right] .  \label{hs2}
\end{eqnarray}%
It is convenient to rescale the Mandelstam variables as follows%
\begin{equation}
s=p_{1}N,\quad t=p_{2}N,\quad u=p_{3}N.  \label{mand1}
\end{equation}%
so that%
\begin{equation*}
p_{1}+p_{2}+p_{3}=1.
\end{equation*}%
In terms of the rescaled variable, the Regge and the hard scattering limits
of the Veneziano amplitude respectively read 
\begin{eqnarray}
&&\left( N\rightarrow \infty ,\left\vert p_{2}\right\vert /\left\vert
p_{1}\right\vert ,\left\vert p_{3}\right\vert /\left\vert p_{1}\right\vert
\quad small\right)   \label{entrop1} \\
A_{V} &\approx &c_{1}\exp \left[ -c_{2}S_{R}^{b}\left( \left\{ p_{i}\right\}
\right) +c_{3}\right] ,  \label{entrop11} \\
&&\left( N\rightarrow \infty ,\left\vert p_{2}\right\vert /\left\vert
p_{1}\right\vert \quad fixed\right)   \label{entrop2} \\
A_{V} &\approx &c_{4}\exp \left[ -N\alpha ^{\prime }S\left( \left\{
p_{i}\right\} \right) +c_{5}\right]   \label{entrop22} \\
c_{1},c_{2},c_{4} &>&0  \notag
\end{eqnarray}%
where $c_{2}$, $c_{3}$, $c_{4}$ and $c_{5}$ are constants and the index $i$
goes from $1$ to $3$. In Eqs. (\ref{entrop1}) and (\ref{entrop11}) it has
been taken into account that $p_{2}$ and $p_{3}$ are small with respect to $%
p_{1}\ $and the index of the Renyi entropy is determined by%
\begin{equation*}
\frac{b}{1-b}=\alpha _{0}(t)\approx 1+o(p_{2})\Rightarrow b\approx \frac{1}{2%
}+o(p_{2}),
\end{equation*}%
where, in this context, it is natural to interpret the constant $N$ as a
measure of the number of degrees of freedom of the system and, consequently,
the limit $N\rightarrow \infty $\ as a thermodynamical limit. Thus, we have
obtained the interesting result that the complex exponential of (minus) the
Renyi entropy can be seen (once the probabilities $p_{i}$ are promoted to
complex variables) as a "Regge-like" limit of the exponential of the
Veneziano amplitude. In other words, (one over) the partition function (one
gets $1/Z$ because in the exponent it appears minus the entropy) of a system
characterized by a Renyi entropy can be seen as the "Regge-limit" of (one
over) the partition function of a system characterized by the standard 
\textit{BGNS}\ entropy in the hard scattering limit. This analysis also
shows that the index of nonextensivity $b$ of the Renyi entropy is related
to the probabilities which, for some reasons, become negligible during the
evolution. To be more specific, the analytic continuation of the exponential
of minus the standard entropy to complex values of the probabilities $p_{i}$
in the Regge limit gives the (complex version of the) exponential of minus
the Renyi entropy with a nonextensive parameter which is connected to the
small probabilities. In this picture, the nonextensivity seems to be related
to states which, in principles, are available but at which, for dynamical
reasons, the probabilities to arrive is negligible. One can look at this
property also from a different perspective: the "statistical" hard
scattering limit is completely symmetric in the probabilities so that
crossing symmetry (which, in this statistical framework, is nothing but the
invariance of the standard entropy under a permutation of the $p_{i}$) is
explicit. On the contrary, the "statistical" Regge limit is not symmetric
with respect to the probabilities (since some are negligible with respect to
others): a breaking of the crossing symmetry leads, upon analytic
continuation, to a nonextensive entropy in which there is a reduced
invariance of the entropy under separate permutations of the large
probabilities and of the small probabilities. Thus, in a sense, the cases in
which, for dynamical reasons, the states (and, consequently, the related
probabilities) are not equivalent\footnote{%
This could happen, for example, when the phase space is made of disconnected
"islands" which, therefore, cannot be compared with continuous
transformations.}, nonextensive entropy should come into play (the present
conclusion is supported by some already known results in the literature:
see, for instance, \cite{El04}).

It is worth to note here that, because of the power of holomorphy, the
requirement of crossing symmetry of the Veneziano amplitude was (together
with the Regge theory) crucial in deriving the explicit expression (\ref%
{vene1}) and its hard scattering and Regge limits. Thus, the high energy
physics analogy suggests that quite in general, once one promotes the
probabilities to complex variables, the analytic continuation of the
exponential of minus the standard entropy very likely is the (complex)
exponential of minus the Renyi entropy (in which the complex probabilities
are the ones with the largest moduli). Also the interpretation of the Regge
poles is quite interesting. The Regge poles of the Veneziano model are
located at%
\begin{equation}
s,t=\alpha ^{\prime }(n-1)  \label{lyconf}
\end{equation}%
so that 
\begin{equation}
\left\vert p_{i}\right\vert =\alpha ^{\prime }\frac{n_{i}-1}{N}
\label{lyconf11}
\end{equation}%
where $n_{i}$ is an integer. Consequently, the analytic continuation of the
exponential of the standard entropy should have poles for negative
probabilities (away from the "physical" real axis) when the moduli of (some
of) the probabilities have a rational values (setting $\alpha ^{\prime }=1$).

The beautiful results of Lee and Yang \cite{LY1,LY2} are very useful in
clarifying the physical meaning of the above analogy between Regge
scattering theory and statistical mechanics. Lee and Yang argued that if one
promotes the inverse temperature $\beta $ and the magnetic field $h$ to be
complex variables (as well as other couplings with external fields which, in
case, could be present) the zeros of the partition function in the complex $%
\beta ,h$ plane lie on lines in the complex planes far away from the real
axis. The number of zeros increases with the size of the system and, in the
thermodynamic limit they should form brunch cuts starting from the real
axis. Hence, from the point of view of complex analysis in the complex $%
\beta ,h$ plane, phase transitions are related to pinching phenomena which
prevent the analytic continuation\footnote{%
The Lee-Yang results have been tested in many situations, however a rigorous
proof in the general case is not available yet.}. It is then quite natural,
instead to promote $\beta $ and $h$ to complex variables while keeping real
the probabilities, to promote the probabilities (and, in case, other local
fields) to be complex variables while keeping real the external parameters.
In this case, the results of Lee and Yang can be restated by saying that the
singularities of the free energy lie on lines in the complex probabilities
plane which, in the thermodynamic limit, tend to form brunch cut emanating
from the real axis. This is precisely what Eqs. (\ref{lyconf}) and (\ref%
{lyconf11}) say: in the thermodynamic limit $N\rightarrow \infty $ the poles
of the Veneziano amplitude (which, as it has been already remarked, is the
inverse of the analytic continuation of a standard partition function)
cumulate forming brunch cut(s) emanating from the physical real axis.

The formal generalization of this relation\ in the cases in which the vector
of probabilities is $N$-dimensional, is a relatively straightforward
computation in string theory in which, instead of computing the vacuum
expectation values on the disk of four tachyonic operators, one considers
the case of $N+1$ tachyonic operators. However, it is worth to stress that
the generalization of the Veneziano amplitude was obtained before the
emergence of its relation with string theory (see, for example, \cite{CC69}%
). In (the exponential of) the standard entropy 
\begin{equation*}
\exp \left[ -\alpha \dsum\limits_{i}^{N}p_{i}\ln p_{i}\right] ,
\end{equation*}%
the probabilities can be thought as complex variables and the above
exponential can be analytically continued in the spirit of Eq. (\ref{mand1})
in which one identifies probabilities with rescaled Mandelstam variables.
The result is a generalized Veneziano amplitude which, in the hard
scattering limit, gives back the (exponential of minus the) standard entropy
and in the Regge limit gives the Renyi behavior (a detailed description of
the analytic structure of the generalized Veneziano amplitude can be found
in \cite{FV69}; an analysis of the mathematical properties of the
generalized Veneziano amplitude in a modern perspective can be found in \cite%
{Kho04}). For instance, one can choose a $k$-dimensional subset $\left\{
p_{1},..,p_{k}\right\} $\ of the probabilities $\left\{ p_{i}\right\} $ with
respect to which to consider the Regge limit\footnote{%
Indeed, it is always possible to arrange the probability vector in such a
way that the probabilities with respect to which one takes the Regge limit
(which are negligible compared to the other probabilities) appear in the
first $k$ components.}:%
\begin{equation*}
N\rightarrow \infty ,\quad \left. \left\{
p_{1},..,p_{k},p_{k+1},..,p_{n}\right\} \right\vert \quad p_{l}\ll
p_{m}\quad \forall l\leq k,\forall m>k.
\end{equation*}%
In this case the Renyi behavior manifests itself as follows%
\begin{equation*}
\exp \left[ -\alpha S\left( k,n-k\right) \right] \approx \exp \left[
-c_{6}S_{R}^{b}\left( P_{i-k}\right) +c_{7}\right] ,
\end{equation*}%
where the notation $S\left( k,n-k\right) $ stresses the fact that one is
taking the Regge limit with respect to the first $k$ probabilities, $P_{i-k}$
is the total probability to be in one of the "non negligible states" (that
is, the states which have non negligible probabilities in the Regge limit), $%
c_{6}$ and $c_{7}$ are suitable constants and the nonextensivity parameter $%
b $ is connected to the negligible probabilities%
\begin{equation*}
b=b(\alpha _{0}(p_{i_{k}})).
\end{equation*}%
It is interesting to note that, as long as none of the probabilities is
negligible with respct to the others, the entropy keeps its standard form
while, when some probabilities are negligible, the nonextensive behavior
manifests itself. The power of holomorphy together with the crossing
symmetry strongly constraint the expected singularities in the complex
probabilities plane. The above observations together with Regge theory could
be useful to extend the Lee-Yang theorems.

The above method opens the possibility to study the analytic continuation
from standard to Renyi entropy in the spirit of the Lee-Yang theorems \cite%
{LY1,LY2}. For instance, in the mean field approach to spin-glass systems,
the \textit{replica method} works in the cases in which, at high enough
temperatures, one can continue a suitable nonextensive free-energy (with
respect to the nonextensive parameter) to obtain the standard free energy.
However, the remarkable discovery of Parisi\footnote{%
Not to be confused with one of the authors.} \cite{Pa79}\ has been that, at
low temperatures the procedure does not work (this is the so-called "replica
symmetry breaking") because of the presence of many metastable states (only
recently \cite{Gu02,GT02,Ta03} it has been possible to prove rigorously that
the Parisi solution is the right one to describe the mean field glassy
phase). A very important quantity in spin glass theory (to provide an
updated list of references is an hopeless task; for two detailed review, see 
\cite{MPV87,CC05}) is the TAP free energy (introduced in \cite{TAP77}) $%
f_{TAP}$ 
\begin{eqnarray}
f_{TAP} &=&-\frac{1}{2}\sum_{i\neq j}J_{ij}m_{i}m_{j}-\sum_{i}h_{i}m_{i}-%
\frac{\beta }{4}\sum_{i\neq j}J_{ij}\left( 1-m_{i}^{2}\right) \left(
1-m_{j}^{2}\right) +  \notag \\
&&+\frac{1}{2\beta }\sum_{i}\left[ \left( 1+m_{i}\right) \log \left( \frac{%
1+m_{i}}{2}\right) +\right.  \notag \\
&&\left. +\left( 1-m_{i}\right) \log \left( \frac{1-m_{i}}{2}\right) \right]
\label{ftap1}
\end{eqnarray}%
where $m_{i}$ is the mean value of the magnetization on the site $i$, $%
J_{ij} $ are the fluctuating spin couplings (it is usually assumed that they
are Gaussian variables with the same mean and variance). In the framework of
mean field theory, the TAP free energy provides a spin glass system above
the glass transition with a detailed description; the TAP free energy $%
f_{TAP}$, to work also below the transition, needs more refined arguments 
\cite{MPV87}. It is interesting to promote, as in the previous case, the
variables $m_{i}$ to complex variables and to try to continue\footnote{%
It is worth to note here that typical tools of high energy physics, such as
the BRST formalism and supersymmetry, already appeared in spin glass theory
(see, for example, \cite{ACG03} and references therein).} the exponential of 
$f_{TAP}$. Since the first tree terms in $f_{TAP}$ are polynomials, their
analytic continuation does not present problems. It is the last term,
related to the (standard \textit{BGNS}) entropy, which generically has a non
trivial analytic continuation. In the high temperature phase, crossing
symmetry in the variables $p_{i}$ defined below, is a good symmetry, so that
previous arguments based on the Veneziano amplitude tell that the analytic
continuation of the entropic term should give rise to a new entropic term in
which, in fact, the Renyi entropy appears. Thus, one can expect that%
\begin{eqnarray}
f_{TAP}^{ac} &\sim &-\frac{1}{2}\sum_{i\neq
j}J_{ij}m_{i}m_{j}-\sum_{i}h_{i}m_{i}-\frac{\beta }{4}\sum_{i\neq
j}J_{ij}\left( 1-m_{i}^{2}\right) \left( 1-m_{j}^{2}\right) +  \notag \\
&&+\frac{\alpha }{2\beta }\log _{2}\left[ \frac{1}{N}\left(
\dsum\limits_{i}\left( p_{i}\right) ^{b}+\left( \overline{p}_{i}\right)
^{b}\right) ^{\frac{1}{1-b}}\right] ,  \label{rsb1} \\
p_{i} &=&\frac{1+m_{i}}{2N},\quad \overline{p}_{i}=\frac{1-m_{i}}{2N},\quad
i,j=1,..,N  \notag
\end{eqnarray}%
where $b$ and $\alpha $ are suitable constants and the last term in the
right hand side of Eq. (\ref{rsb1}) is the \textit{Reny}-like term. A
comparison with known results in spin glass theory the \cite{KPV93} \cite%
{CK93} \cite{CS95} shows that the interpretation of the standard \textit{BGNS%
} entropic term in the TAP free energy as the hard scattering limit of a
suitable string theoretical amplitude gives, in the Regge limit, an
expression which bears a strong resemblance with the low temperature TAP
free energy suitable to study the glassy phase. Thus, the hard scattering
and Regge limits correspond to unbroken replica phase and replica symmetry
breaking respectively. The non trivial benefit of this method is that,
besides to clarify the physical origin of "nonextensivity", it could be easy
to handle because of the huge amount of works (in which one could find
explicit computations important for statistical mechanics) already available
in string theory.

\end{document}